\documentclass[prb,aps,twocolumn,amsmath,amssymb,floatfix,showpacs,
superscriptaddress]{revtex4}

\usepackage[dvips]{graphics}
\usepackage{color}
\definecolor{dred}{rgb}{0,0,0.6}

\begin{document}

\title{\textcolor{dred}{Anomalous magnetic response of a quasi-periodic 
mesoscopic ring in presence of Rashba and Dresselhaus spin-orbit 
interactions}}

\author{Moumita Patra}

\affiliation{Physics and Applied Mathematics Unit, Indian Statistical
Institute, 203 Barrackpore Trunk Road, Kolkata-700 108, India}

\author{Santanu K. Maiti}

\email{santanu.maiti@isical.ac.in}

\affiliation{Physics and Applied Mathematics Unit, Indian Statistical
Institute, 203 Barrackpore Trunk Road, Kolkata-700 108, India}

\begin{abstract}

We investigate the properties of persistent charge current driven by 
magnetic flux in a quasi-periodic mesoscopic Fibonacci ring with Rashba
and Dresselhaus spin-orbit interactions. Within a tight-binding framework
we work out individual state currents together with net current based on
second-quantized approach. A significant enhancement of current is observed
in presence of spin-orbit coupling and sometimes it becomes orders of
magnitude higher compared to the spin-orbit interaction free Fibonacci 
ring. We also 
establish a scaling relation of persistent current with ring size, 
associated with the Fibonacci generation, from which one can directly 
estimate current for any arbitrary flux, even in presence of spin-orbit 
interaction, without doing numerical simulation. The present analysis 
indeed gives a unique opportunity of determining persistent current and 
has not been discussed so far.

\end{abstract}


\maketitle

\section{Introduction}

Over the last couple of decades the phenomenon of persistent charge 
current in mesoscopic ring structures has drawn a lot of attention due 
to its crucial role in understanding quantum coherence in such 
interferometric geometries. In the early $80$'s B\"{u}ttiker {\em et al.}
first proposed theoretically~\cite{butt1} that a small conducting ring 
carries a net circulating charge current in presence of magnetic flux 
$\phi$. This is a pure quantum mechanical phenomenon and can sustain 
even in presence of disorder. The experimental verification of persistent 
charge current came into realization during $1990$ through the significant 
experiment~\cite{levy} done by Levy {\em et al.} considering $10^7$ 
isolated mesoscopic copper rings. Later many experimental verifications 
and theoretical propositions have been made~\cite{jari,bir,chand,blu,
ambe,schm1,schm2,peet,spl,ding} towards this direction.

A large part of the literature reported so far describes the phenomenon 
of persistent currents considering perfect periodic rings as well as 
completely random ones~\cite{gefen,smt,mont,ph,san1}. But a little less 
attention 
was paid to the quasi-periodic ring structures~\cite{fib1,fib2,fib3,fib4} 
which actually bridge the gap between these fully ordered and randomly 
disordered phases. However, the studies involving persistent current in 
quasi-periodic ring geometries are mostly confined within non-interacting 
picture and, to the best of our knowledge, no one has addressed its behavior 
in presence of spin-orbit (SO) interaction which can bring significant 
new features into light. It is therefore worthwhile to analyze the 
characteristics of persistent current in a quasi-periodic Fibonacci ring 
considering the effect of spin-orbit interaction (SOI).

Usually two different types of SO 
interactions~\cite{rashba,dressel,winkler,sheng}, 
namely Rashba and Dresselhaus, are encountered in solid state materials 
depending on their sources. The Rashba SO coupling is originated by 
breaking the inversion symmetry of the structure, which can be thus tuned 
via external gate electrodes~\cite{meier} placed in the vicinity of the 
sample. While, the other SO coupling cannot be controlled by external 
means as it is generated from the bulk inversion asymmetry.

In the present paper we make a comprehensive analysis of non-decaying 
circular current in a quasi-periodic mesoscopic Fibonacci ring subjected 
\begin{figure}[ht]
{\centering \resizebox*{5.8cm}{3.5cm}{\includegraphics{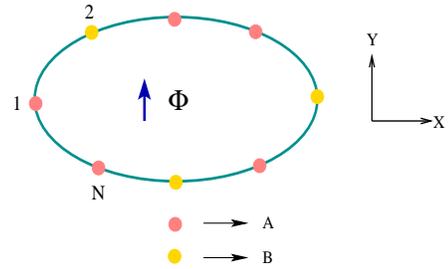}}\par}
\caption{(Color online). Schematic diagram of a $5$th generation 
quasi-periodic Fibonacci mesoscopic ring subjected to Rashba and 
Dresselhaus spin-orbit interactions. The ring, composed of two different
types of atomic sites A and B those are represented by two distinct 
colored filled circles, carries a net circulating charge current in 
presence of magnetic flux $\phi$.}
\label{FF}
\end{figure}
to Rashba and Dresselhaus SO couplings. Two primary lattices, 
viz, $A$ and $B$ are used to get a $N$-site Fibonacci chain following the 
generation rule $F_m$ ($m \geq 3$) $=\left\{F_{m-1}, F_{m-2}\right\}$ with 
$F_1=A$ and $F_2=AB$, which is then bent and coupled at its two ends 
to form a ring. Alternatively, we can think that $m$th generation Fibonacci
sequence $F_m$ can be constructed using two lattice sites $A$ and $B$ applying
$m$ times the inflation rules $A \rightarrow AB$ and $B \rightarrow A$
recursively, starting with the lattice $A$ or $B$. Here we start with the
lattice $A$, for the sake of simplicity, and thus, $A$, $AB$, $ABA$, $ABAAB$,
$ABAABABA$, $\dots$, etc., are the first few generations of the Fibonacci
sequence. Therefore, as an example, $F_4=ABAAB$ forms a $5$-site ($N=5$) 
Fibonacci ring. This is one representation, the so-called site 
model~\cite{fib2,fib4}, of a Fibonacci generation. Another form of it 
is also conveniently used which is known as bond model~\cite{bond1,bond2} 
where long ($L$) and short ($S$) bonds are taken into account, setting 
identical lattice sites. In few cases mixed model~\cite{fib4}, a 
combination of site and bond models, is also used in studying electronic 
behavior. For the sake of simplicity here we restrict ourselves to the 
first configuration. 

Based on a tight-binding (TB) framework we compute persistent current
using second-quantized approach~\cite{san2}. With this formalism one 
can find current carried by individual energy levels, and, from that total 
current for
a particular band filling can be easily estimated. The major advantage of
this technique is that, it reduces numerical errors especially for larger
rings by avoiding the derivative of ground state energy with respect to 
flux $\phi$, as used in conventional current 
calculations~\cite{gefen,spl,san3}. Most importantly, studying individual 
state currents conducting nature of different eigenstates can be 
determined which is quite significant to understand the response of
a complete system. Thus, utilizing it, the crucial role played by SO 
interactions on current carrying states can be analyzed clearly, which is
one key motivation behind this work. We find that state currents get
increased significantly with SO coupling, which thus provide a large net
current and sometimes it becomes orders of magnitude higher than the
SOI-free Fibonacci rings. Undoubtedly this is an important
observation and might throw some light in the era of deep-rooted debate 
between the experimental observations and theoretical estimates of current 
amplitudes.

Apart from this, we also discuss the behavior of persistent current for 
different band fillings, and, on its won merit, the quasi-periodic 
structure exhibits several anomalous features which can have great 
signature, particularly, in the aspect of controlling conducting nature
of the full system.

Finally, we make a detailed analysis to find a scaling relation of 
persistent current with ring size $N$, associated with the generation
$F_m$. From our extensive numerical analysis we establish that for a 
typical flux $\phi$, the current obeys a relation $C N^{-\xi}$, where
$\xi$ depends on the ratio between the site energy difference and 
nearest-neighbor hopping integral. Thus keeping the ratio constant, 
site energies as well as hopping integral can be tuned and with these 
changes $\xi$ remains invariant. The pre-factor $C$ strongly depends 
on both SO coupling and magnetic flux, which is also reported here in 
detail for the completeness. These results offer a unique 
opportunity to determine persistent current in a Fibonacci ring,
subjected to SO coupling, for any arbitrary flux $\phi$ without
doing any numerical simulation. This is another essential motivation
for the present investigation.

We organize the rest of the article as follows. In Sec. II we present
the model and its Hamiltonian in tight-binding framework. The procedure
for calculating persistent current carried by different eigenstates as
well as the net current for a particular electron filling is given in
Sec. III, and the numerical results are discussed in Sec. IV. Finally,
in Sec. V we summarize our main results.

\section{Model and Tight-binding Hamiltonian}

We start by referring to Fig.~\ref{FF}, where a quasi-periodic mesoscopic 
Fibonacci ring composed of two different types of atomic sites $A$ and $B$
is given. The ring, subjected to both Rashba and Dresselhaus SO interactions,
carries a net circulating charge current in presence of an AB flux $\phi$.

To illustrate this model quantum system we adopt a tight-binding framework.
In the absence of electron-electron interaction the TB Hamiltonian for a
$N$-site Fibonacci ring can be described as following:
\begin{equation}
H = H_{\mbox{\tiny 0}} + H_{\mbox{\tiny rashba}} + H_{\mbox{\tiny dressl}}. 
\label{equ1}
\end{equation}
The first term, $H_{\mbox{\tiny 0}}$, represents the Fibonacci ring in the
absence of SO interactions and it becomes
\begin{equation}
H_{\mbox{\tiny 0}}=\sum_n \mbox{\boldmath $c$}_n^{\dag} 
\mbox{\boldmath $\epsilon$} \mbox{\boldmath $c$}_n  
+ \sum_n \left(e^{i\theta}\mbox{\boldmath $c$}_{n+1}^{\dag} 
\mbox{\boldmath $t$} \mbox{\boldmath $c$}_n + 
e^{-i\theta}\mbox{\boldmath $c$}_{n}^{\dag} 
\mbox{\boldmath $t$}^{\dag} \mbox{\boldmath $c$}_{n+1} \right)
\label{equ2}
\end{equation}
where $\theta=2 \pi \phi/N$ is the phase factor due to the flux $\phi$
which is measured in unit of the elementary flux quantum $\phi_0$ ($=ch/e$),
and $n=1$, $2$, $3$ $\dots$. The other factors are described as follows.\\
$\mbox{\boldmath $c$}_n=\left(\begin{array}{c}
c_{n\uparrow} \\
c_{n\downarrow}\end{array}\right)$ and 
$\mbox{\boldmath $c$}^{\dagger}_n=\left(\begin{array}{cc}
c_{n\uparrow}^{\dagger} & c_{n\downarrow}^{\dagger} 
\end{array}\right)$, where $c_{n\sigma}^{\dagger}$
($c_{n\sigma}$) is the creation (annihilation) operator 
for an electron at $n$-th site with spin $\sigma(\uparrow,\downarrow)$. 
Considering the on-site potential at $n$th site for an electron with
spin $\sigma$ as $\epsilon_{n\sigma}$ we express
$\mbox{\boldmath $\epsilon$}_n=\left(\begin{array}{cc}
\epsilon_{n\uparrow} & 0 \\
0 & \epsilon_{n\downarrow} \end{array}\right)$. Depending on the atomic 
site $A$ or $B$, $\epsilon_{n\sigma}$ becomes $\epsilon_{n\sigma}^A$ or
$\epsilon_{n\sigma}^B$. $\mbox{\boldmath $t$}$ is ($2\times 2$) diagonal 
matrix with the diagonal elements 
$\mbox{\boldmath $t$}_{11}=\mbox{\boldmath $t$}_{22}=t$, where $t$ represents
the nearest-neighbor hopping integral.

The second term, $H_{\mbox{\tiny rashba}}$, describes the Hamiltonian 
associated with Rashba SO coupling~\cite{san2} and it becomes
\begin{eqnarray}
H_{\mbox{\tiny rashba}}&=&-\sum_n\alpha\left[\mbox{\boldmath $c$}_{n+1}^{\dag} 
\left(i\mbox{\boldmath $\sigma$}_x \cos\varphi_{n,n+1} + 
i\mbox{\boldmath $\sigma$}_y \sin\varphi_{n,n+1}\right)\right.\nonumber \\
& & \left. e^{i\theta} 
\mbox{\boldmath $c$}_n + h.c. \right]
\label{equ3}
\end{eqnarray}
where $\alpha$ measures the Rashba SO coupling strength and 
$\varphi_{n,n+1}=\left(\varphi_n + \varphi_{n+1}\right)/2$ with 
$\varphi_n=2\pi(n-1)/N$. $\mbox{\boldmath $\sigma$}_x$ and
$\mbox{\boldmath $\sigma$}_y$ are the Pauli spin matrices in 
$\mbox{\boldmath $\sigma$}_z$ diagonal representation.

In a quite similar way we can write the last term of the total Hamiltonian
Eq.~\ref{equ1} which is related to Dresselhaus SO coupling~\cite{san2} as,
\begin{eqnarray}
H_{\mbox{\tiny dressl}}&=&\sum_n\beta\left[\mbox{\boldmath $c$}_{n+1}^{\dag} 
\left(i\mbox{\boldmath $\sigma$}_y \cos\varphi_{n,n+1} + 
i\mbox{\boldmath $\sigma$}_x \sin\varphi_{n,n+1}\right)\right.\nonumber \\
& & \left. e^{i\theta} 
\mbox{\boldmath $c$}_n + h.c. \right]
\label{equ4}
\end{eqnarray}
where $\beta$ is the Dresselhaus coefficient.

\section{Theoretical Formulation}

In this section, we calculate persistent charge current carried by 
individual eigenstates using the second-quantized approach and from 
these individual state currents we determine the net current for a 
particular electron filling.

We start with the current operator $\mbox{\boldmath$I$}=e 
\mbox{\boldmath$\dot{x}$}/(Na)$, where $a$ is the lattice spacing and 
\mbox{\boldmath$\dot{x}$} is the velocity operator written in the form,
\begin{equation}
\mbox{\boldmath$\dot{x}$} = \frac{1}{i\hbar}
\left[\mbox{\boldmath$x$},\mbox{\boldmath$H$}\right]
\label{equ5}
\end{equation}
where $\mbox{\boldmath$x$}=a\sum\limits_n \mbox{\boldmath$c$}_n^{\dagger}
n \mbox{\boldmath$c$}_n$ denotes the position operator. Thus we can write
the current operator as
\begin{equation}
\mbox{\boldmath$I$}=\frac{e}{Na}\frac{1}{i\hbar}\left[\mbox{\boldmath$x$},
\mbox{\boldmath$H$}\right]=\frac{2\pi i e}{N a h}\left[\mbox{\boldmath$H$},
\mbox{\boldmath$x$}\right].
\label{equ6}
\end{equation}
Substituting \mbox{\boldmath$x$} and \mbox{\boldmath$H$} into Eq.~\ref{equ6}
and doing quite lengthy but straightforward calculations we eventually reach
to the expression
\begin{equation}
\mbox{\boldmath$I$}=\frac{2\pi i e}{Nh}\sum\limits_n
\left(\mbox{\boldmath$c$}_n^{\dagger} 
\mbox{\boldmath$t$}_{\varphi}^{\dagger n,n+1} 
\mbox{\boldmath$c$}_{n+1} e^{-i\theta} - \mbox{\boldmath$c$}_{n+1}^{\dagger}
\mbox{\boldmath$t$}_{\varphi}^{n,n+1} \mbox{\boldmath$c$}_n e^{i\theta}
\right)
\label{equ7}
\end{equation}
where $\mbox{\boldmath$t$}_{\varphi}^{n,n+1}$ is a ($2\times 2$) matrix 
whose elements are as follows:
$\mbox{\boldmath$t$}_{\varphi,11}^{n,n+1}=
\mbox{\boldmath$t$}_{\varphi,22}^{n,n+1}=t$, 
$\mbox{\boldmath$t$}_{\varphi,12}^{n,n+1}=-i\alpha e^{-i\varphi_{n,n+1}}
+ \beta e^{i\varphi_{n,n+1}}$,
$\mbox{\boldmath$t$}_{\varphi,21}^{n,n+1}=-i\alpha e^{i\varphi_{n,n+1}}
- \beta e^{i\varphi_{n,n+1}}$.
\vskip 0.2cm
\noindent
Once \mbox{\boldmath$I$} is established, the current carried by any energy
eigenstate $|\psi_m\rangle$ (say) can be calculated by the relation
\begin{equation}
I_m=\langle\psi_m|\mbox{\boldmath$I$}|\psi_m\rangle 
\label{equ8}
\end{equation}
where $|\psi_m\rangle=\sum\limits_n\left(a_{n\uparrow}^m
|n\uparrow \rangle + a_{n\downarrow}^m |n\downarrow \rangle\right)$.
$|n\sigma\rangle$'s are the Wannier states and $a_{n\sigma}^m$'s are 
the coefficients. After simplification we reach to the following
\begin{eqnarray}
I_m & = & \frac{2\pi ie}{Nh}\sum\limits_{n}\left(t a_{n,\uparrow}^{m\ast}
a_{n+1,\uparrow}^m e^{-i\theta}-t a_{n+1,\uparrow}^{m\ast}
a_{n,\uparrow}^m e^{i\theta}\right)\nonumber \\
 & & +\frac{2\pi ie}{Nh}\sum\limits_{n}\left(t a_{n,\downarrow}^{m\ast}
a_{n+1,\downarrow}^m e^{-i\theta}-t a_{n+1,\downarrow}^{m\ast}
a_{n,\downarrow}^m e^{i\theta}\right)\nonumber \\
& & +\frac{2\pi ie}{Nh}\sum\limits_{n}\left\{\left(i\alpha
e^{-i\phi_{n,n+1}}+\beta e^{i\phi_{n,n+1}}\right) \times \right.\nonumber \\
& & \left. a_{n,\uparrow}^{m\ast} a_{n+1,\downarrow}^m e^{-i\theta} 
\right.\nonumber \\
& & +\left.\left(i\alpha e^{i\phi_{n,n+1}}+\beta
e^{-i\phi_{n,n+1}}\right)a_{n+1,\downarrow}^{m\ast} a_{n,\uparrow}^m
e^{i\theta}\right\}\nonumber\\
& & +\frac{2\pi ie}{Nh}\sum\limits_{n}\left\{\left(i\alpha
e^{i\phi_{n,n+1}}+\beta e^{-i\phi_{n,n+1}}\right)\times \right.\nonumber \\
 & & \left. a_{n,\downarrow}^{m\ast} a_{n+1,\uparrow}^m e^{-i\theta}
\right.\nonumber \\
& & +\left.\left(i\alpha e^{-i\phi_{n,n+1}}-\beta
e^{i\phi_{n,n+1}}\right)a_{n+1,\uparrow}^{m\ast}
a_{n,\downarrow}^m e^{i\theta}\right\}
\label{equ9}
\end{eqnarray}
This is the general expression of persistent charge current carried by 
an eigenstate $|\psi_m \rangle$ in presence of Rashba and Dresselhaus
SO interactions. With this relation total charge current at absolute 
zero temperature ($T=0\,$K) for a $N_e$-electron system becomes
\begin{equation}
I=\sum\limits_{m=1}^{N_e} I_m
\label{equ10}
\end{equation}
where the contributions from the lowest $N_e$ states are taken into account.

This is one way (viz, the second-quantized approach) of calculating 
persistent charge current which we use in this work due its potentiality 
for our present analysis. But, there exists another method, the so-called
derivative method~\cite{gefen,spl}, where net circulating current is 
evaluated by taking a first order derivative of ground state energy 
$E_0$ (say) with respect to AB flux $\phi$.

\section{Numerical Results and Discussion}

According to the theoretical formulation introduced in Sec. III we are
now ready to analyze numerical results, computed in the limit of zero
temperature, for charge current carried by individual energy levels,
net current for a particular electron filling and its scaling behavior 
with system size in presence of Rashba and Dresselhaus SO interactions.
In our model since the sites are non-magnetic we can write 
$\epsilon_{n\sigma}^A$ simply as $\epsilon_A$ for all $A$-type atomic 
sites, and similarly, for $B$-type sites $\epsilon_{n\sigma}^B=\epsilon_B$.
When $\epsilon_A=\epsilon_B$, the system becomes a perfect ring as 
on-site energies are independent of site index $n$, and thus, we can set
them to zero without loss of any generality. All the energies used in our
calculations are scaled with respect to the nearest-neighbor hopping
integral $t$ which is fixed at $1\,$eV throughout the presentation, and,
we measure the current in unit of $et/h$.

Before addressing the central results of persistent 
current, let us have a look at the energy band spectrum for both perfectly
ordered and Fibonacci rings to make the present work a self contained one.

\subsection{Energy Spectrum}

\begin{figure}[ht]
{\centering \resizebox*{8cm}{7cm}{\includegraphics{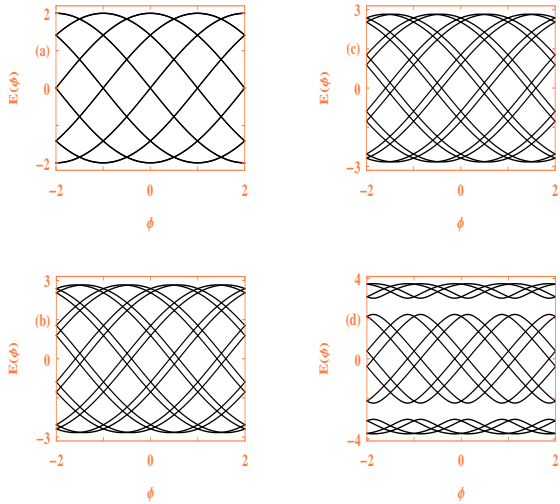}}\par}
\caption{(Color online). Electronic energy levels of a 
$8$-site ring ($5$th generation) with $\epsilon_A=\epsilon_B=0$ as a 
function of flux $\phi$ for different values of $\alpha$ and $\beta$, 
where (a) $\alpha=\beta=0$; (b) $\alpha=1\;$eV, $\beta=0$; (c) $\alpha=0$, 
$\beta=1\;$eV and (d) $\alpha=0.5\;$eV, $\beta=1.5\;$eV.}
\label{pfs}
\end{figure}

\begin{figure}[ht]
{\centering \resizebox*{8cm}{7cm}{\includegraphics{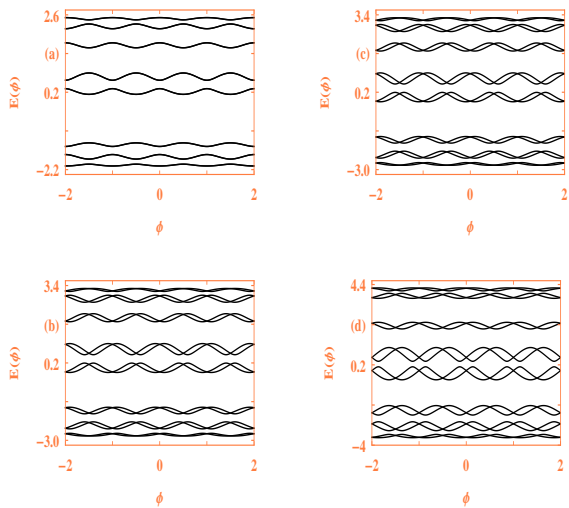}}\par}
\caption{(Color online). Electronic energy levels of a
$8$-site ring ($5$th generation) with $\epsilon_A=-\epsilon_B=1\;$eV as 
a function of flux $\phi$ for different values of $\alpha$ and $\beta$, 
where (a), (b), (c) and (d) correspond to the identical meaning as given
in Fig.~\ref{pfs}.}
\label{fs}
\end{figure}

In Figs.~\ref{pfs} and \ref{fs} flux dependent energy spectra 
are shown for a $8$-site perfectly ordered (viz, $\epsilon_A=\epsilon_B=0$) 
and Fibonacci ($\epsilon_A=-\epsilon_B=1\;$eV) rings, respectively. From the
spectra it is clearly observed that the correlated disorder removes the 
energy level crossings noticed in the perfect case and also it reduces
the slope of the energy levels. Most importantly we see that the number of
energy levels gets twice when the ring is subjected to both AB flux $\phi$ 
and SO interaction compared to the SOI-free ring. From the fundamental
principle of quantum mechanics it is well known that if the Hamiltonian
is symmetric under time-reversal operation the Kramer's degeneracy gets
preserved, resulting degenerate energy levels. For our model, the two
physical parameters, magnetic flux and SO coupling, affect the degeneracy.
In presence of $\phi$ two-fold degenerate energy levels 
are obtained
from the SOI-free (viz, $\alpha=\beta=0$) ring. Similar kind of
two-fold degenerate energy states are also noted under time-reversal
symmetry condition (i.e., $\phi=0$) when the ring is subjected to SO
coupling. For this situation we can write $E(k,\uparrow)=E(-k,\downarrow)$
following the Kramer's degeneracy, where $k$ represents the wave vector. But,
it disappears completely as long as the magnetic flux is introduced
($E(k+\phi,\uparrow) \ne E(-k+\phi,\downarrow)$), and therefore, we get
twice distinct energy levels compared to the SOI-free AB ring.
\begin{figure}[ht]
{\centering \resizebox*{8cm}{7cm}
{\includegraphics{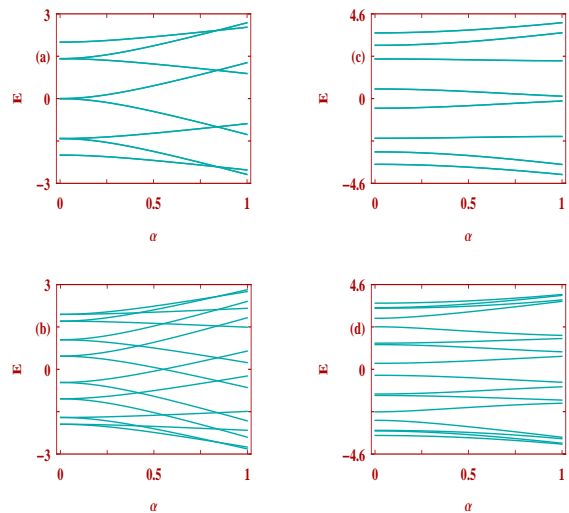}}\par}
\caption{(Color online). Electronic energy levels as a 
function of Rashba SO coupling $\alpha$ of a $8$-site ring ($5$th 
generation) with $\epsilon_A=\epsilon_B=0$ for different values of 
$\phi$ and $\beta$, where (a) $\phi=\beta=0$; (b) $\phi=0.3$, $\beta=0$; 
(c) $\phi=0$, $\beta=1.5\;$eV and (d) $\phi=0.3$, $\beta=1.5\;$eV.}
\label{pfsa}
\end{figure}
\begin{figure}[ht]
{\centering \resizebox*{8cm}{7cm}
{\includegraphics{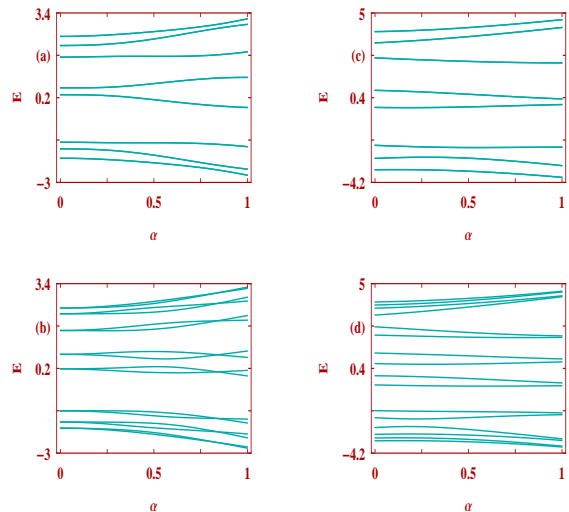}}\par}
\caption{(Color online). Electronic energy levels as a
function of Rashba SO coupling $\alpha$ of a $8$-site ring ($5$th
generation) with $\epsilon_A=-\epsilon_B=1\;$eV for different values of
$\phi$ and $\beta$, where (a), (b), (c) and (d) correspond to the 
identical meaning as given in Fig.~\ref{pfsa}.}
\label{fsa}
\end{figure}
In addition it is also crucial to note that even for 
perfectly ordered ring finite gaps appear near the two edges of the energy
band spectrum when both the Rashba and Dresselhaus SO interactions
are present (see Fig.~\ref{pfs}(d)). The origin of such gaps in a ring
system with $\alpha$ and $\beta$ has been described elaborately by
Chang {\em et al}~\cite{sheng} in $2006$ and they have shown how the
gap is sensitive with these parameter values.

In order to understand the precise role of SO coupling
on energy levels in Figs.~\ref{pfsa} and \ref{fsa} we present the SO 
coupling dependent spectra for perfectly ordered
and Fibonacci rings, respectively, considering the identical ring size
as taken in Figs.~\ref{pfs} and \ref{fs}, for different values of $\phi$
and $\beta$. With increasing the SO interaction strength splitting of the
energy levels gets wider, while the degeneracy factors in different 
diagrams remains identical as discussed in the spectra Figs.~\ref{pfs} 
and \ref{fs}. In these SOI dependent spectra (Figs.~\ref{pfsa} and 
\ref{fsa}) eigenenergies are plotted as function of Rashba SO coupling
setting some typical values of $\beta$. Exactly similar feature is also
obtained under swapping the parameters $\alpha$ and $\beta$ (not shown
here to save space), and its origin can be understood from the forthcoming 
sub-section.

\subsection{Enhancement of persistent current}

Let us start with discussing the influence of SO couplings on the behavior 
of persistent current carried by individual energy eigenstates for a 
\begin{figure}[ht]
{\centering \resizebox*{8cm}{10cm}{\includegraphics{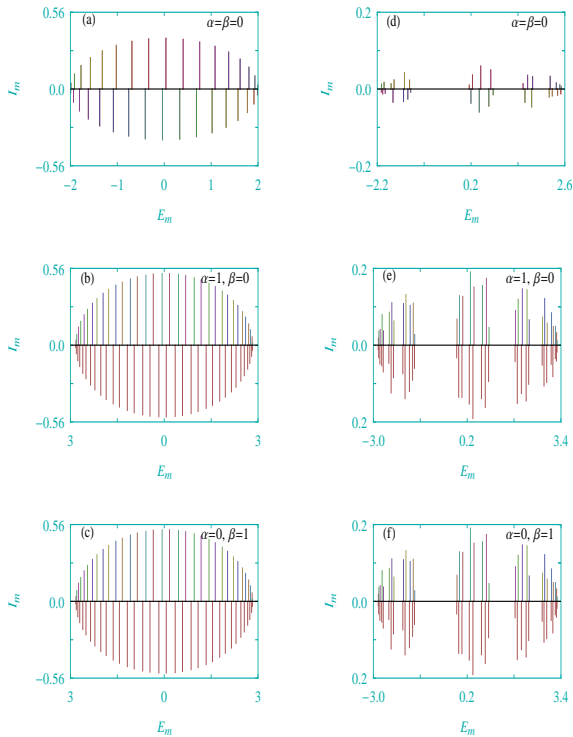}}\par}
\caption{(Color online). Variation of persistent current $I_m$ carried by 
individual energy eigenstates $|\psi_m\rangle$ having eigenenergy $E_m$
of a $8$th generation Fibonacci ring at the typical flux $\phi=0.4$ for 
different values of Rashba and Dresselhaus SO interactions. In the left 
column we set $\epsilon_A=\epsilon_B=0$, while in the right column we choose
$\epsilon_A=-\epsilon_B=1\;$eV.}
\label{CCN}
\end{figure}
typical flux $\phi$. The results of a $8$th generation Fibonacci ring 
are shown in 
Fig.~\ref{CCN} considering $\phi=0.4$, where the left column corresponds 
to $\epsilon_A=\epsilon_B=0$, while for the right column we choose
$\epsilon_A=-\epsilon_B=1$. Several interesting features are obtained 
those are analyzed as follows. 

At a first glance one can see that in the absence of SO coupling all 
distinct energy levels carry finite currents for the perfectly ordered ring
\begin{figure}[ht]
{\centering \resizebox*{7cm}{7cm}{\includegraphics{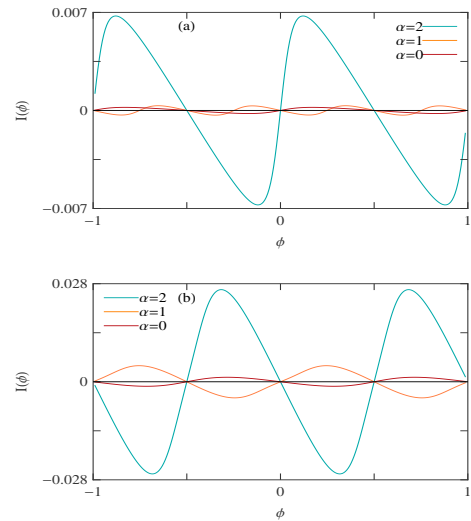}}\par}
\caption{(Color online). Net current at a particular electron filling as a
function of flux $\phi$ for some Fibonacci rings 
($\epsilon_A=-\epsilon_B=1\;$eV) with different values of $\alpha$, where 
the red, orange and cyan curves correspond to $\alpha=0$, $1$ and $2\;$eV,
respectively. The other physical parameters are: (a) $N=610$ ($14$th 
generation), $N_e=400$ and (b) $N=377$ ($13$th generation), $N_e=300$. 
Here we set $\beta=0$.}
\label{CF}
\end{figure}
($\epsilon_A=\epsilon_B=0$), whereas these currents almost cease to zero
in the case of correlated disordered ring ($\epsilon_A=-\epsilon_B=1\;$eV).
This is quite obvious since a pure ring provides extended states which
carry finite currents, while almost localized states obtained from the 
Fibonacci ring yield vanishingly small currents. These currents even more
decrease with increasing the correlation strength 
($|\epsilon_A \sim \epsilon_B|$) (which are not shown here in the figure).
This fact has already been discussed in literature in connection with the
localization aspects of different aperiodic crystal classes. But one of the 
major issues of our present investigation i.e., the interplay between 
SO interactions and quasiperiodic Fibonacci sequence on electronic 
localization has not been addressed earlier. To illustrate it, in the middle
and last rows of Fig.~\ref{CCN} we show the dependence of state currents 
on $\alpha$ and $\beta$, respectively. A large number of discrete states
of the Fibonacci ring, those were almost localized in absence of SO 
coupling (Fig.~\ref{CCN}(d)), provides sufficiently large current in 
presence of non-zero SO coupling. This enhancement of current in presence
of SO coupling can be elucidated in terms of quantum interference as it
is directly related to the localization process. In presence of disorder,
quantum interference gets dominated which gives rise to the electronic
localized states, while this effect becomes weakened as a results of SO
coupling as it involves spin-flipping, resulting enhanced charge currents.
Naturally, this effect will be reflected into the net current for a 
particular band filling as discussed later. In addition, it is important to
\begin{figure}[ht]
{\centering \resizebox*{7cm}{7cm}
{\includegraphics{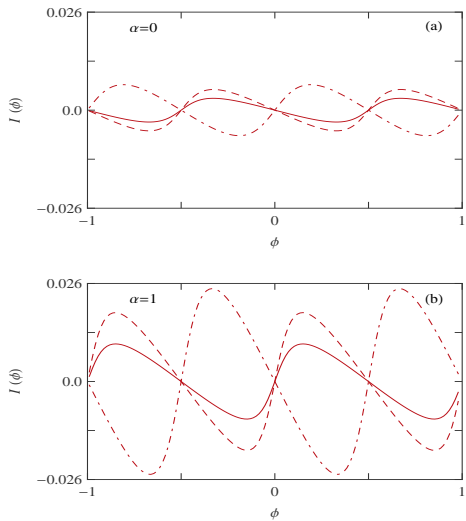}}\par}
\caption{(Color online). Filling dependent current-flux characteristics of
a $11$th generation Fibonacci ring ($\epsilon_A=-\epsilon_B=1\;$eV) where
(a) $\alpha=0$ and (b) $\alpha=1\;$eV. The solid, dashed and dot-dashed 
curves correspond to $N_e=18$, $34$, and $56$, respectively. The Dresselhaus
SO coupling is fixed at zero.}
\label{FD}
\end{figure}
note that though the perfect ring exhibits extended states, they even 
carry higher currents in presence of finite SO coupling which is clearly
spotted from the spectra given in the left column of Fig.~\ref{CCN}.

Figure~\ref{CCN} also depicts that the nature (viz, magnitude and phase) 
of current carrying states remain unchanged under swapping the parameters 
$\alpha$ and $\beta$. This invariant nature can be understood through a 
simple mathematical argument. Inspecting carefully the Rashba and Dresselhaus
Hamiltonians one can see that they are connected by a unitary transformation
$U^{\dagger} H_{\mbox{\tiny rashba}} U = H_{\mbox{\tiny dressl}}$, where
$U=(\mbox{\boldmath$\sigma$}_x + \mbox{\boldmath$\sigma$}_y)/\sqrt{2}$ is
the unitary matrix. Therefore, any eigenstate $|\psi_p \rangle$ (say) of 
the Rashba ring can be
written in terms of the eigenstate $|\psi_p^{\prime} \rangle$ of the 
Dresselhaus ring where $|\psi_p \rangle = U |\psi_p^{\prime} \rangle$.
This immediately gives the current for the Dresselhaus ring:
$I_p (\mbox{for}~ H_{\mbox{\tiny dressl}})= \langle \psi_p^{\prime}|
\mbox{\boldmath$I$}|\psi_p^{\prime} \rangle = \langle \psi_p|
U^{\dagger}\mbox{\boldmath$I$}U|\psi_p \rangle = \langle \psi_p|
\mbox{\boldmath$I$}|\psi_p \rangle = 
I_p (\mbox{for}~ H_{\mbox{\tiny rashba}})$.
Hence, it is clearly observed that the nature of the current carrying states 
for the Rashba ring is exactly identical to that of the Dresselhaus ring.
Using $SU(2)$ spin rotation transformation mechanism Sheng 
and Chang~\cite{sheng} have established that the Hamiltonian of the Rashba 
SOI alone is mathematically equivalent to that of the Dresselhaus SOI
alone, and thus, our findings regarding the invariant nature of current 
carrying states under swapping the SOIs are consistent with their analysis.

Following the above characteristics of different current carrying states
now we discuss the behavior of net current for a particular electron
filling.
\begin{figure}[ht]
{\centering \resizebox*{7cm}{7cm}{\includegraphics
{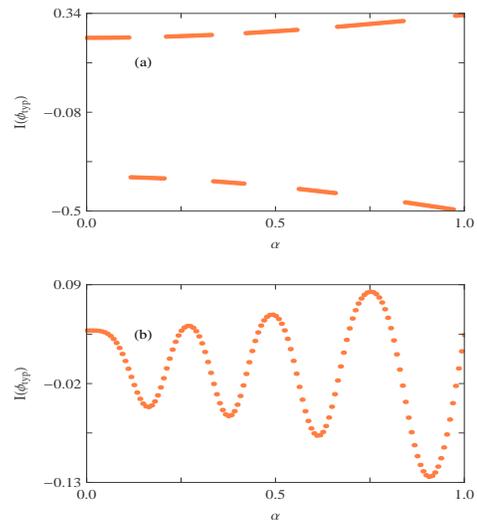}}\par}
\caption{Current, evaluated at a particular flux $\phi=0.3\phi_0$, as a 
function of $\alpha$ ($\beta$ is fixed at zero) for (a) ordered and (b)
Fibonacci ($\epsilon_A=-\epsilon_B=1\;$eV) rings considering $N=34$ 
($8$th generation) and $N_e=20$.} 
\label{CCR}
\end{figure}
\begin{figure}[ht]
{\centering \resizebox*{6.5cm}{7cm}{\includegraphics{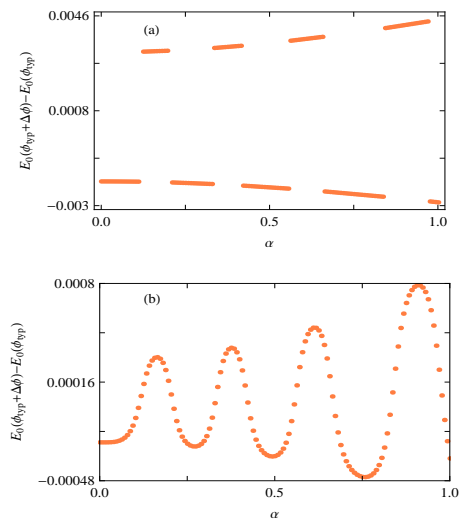}}\par}
\caption{(Color online). Variation of $\Delta E_0$ with $\alpha$ at 
$\phi_{\mbox{\tiny typ}}=0.3\phi_0$ for the same parameter values as taken
in Fig.~\ref{CCR}, where (a) and (b) represent the identical meaning as 
given in Fig.~\ref{CCR}. We choose $\Delta \phi=0.125/16$.}
\label{GER}
\end{figure}
The results are shown in Fig.~\ref{CF}, where we show the 
variation of net persistent current as a function of $\phi$ for some
typical Fibonacci rings considering different values of Rashba SO 
coupling. In (a) the currents are computed for $N=610$ ($14$th generation) 
and $N_e=400$, while in (b) these are performed for $N=377$ ($13$th
generation) and $N_e=300$. From the 
spectra it is observed that the current almost vanishes for the entire
flux window when the ring is free from SO coupling (red curves). This 
is solely due to the aperiodic nature of the site potentials. Introducing
the SO coupling one can achieve higher current (orange lines), and, for a
moderate SO coupling a dramatic change is observed (cyan curves), 
reflecting the $I_m$-$E_m$ spectra given in the right column of 
Fig.~\ref{CCN}.

\subsection{Effect of electron filling}

To test the dependence of persistent current on electron filling, in
Fig.~\ref{FD} we show the current-flux characteristics of a $11$th 
\begin{figure}[ht]
{\centering \resizebox*{6.5cm}{7cm}{\includegraphics{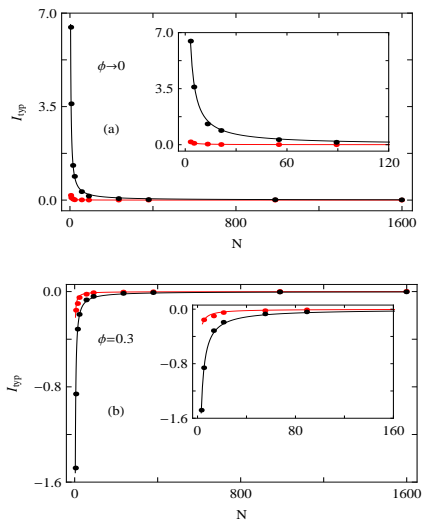}}\par}
\caption{(Color online). Variation of current with ring size $N$ considering
$\epsilon_A=-\epsilon_B=0.5\;$eV and $\beta=0$ where (a) $\phi \rightarrow 0$ 
and (b) $\phi=0.3$. The red and black dots, corresponding to $\alpha=0$ and 
$1.5\;$eV, respectively, are determined from the second-quantized approach.
Using these dots we find scaling relation between $I_{\mbox{\tiny typ}}$
and $N$ which produces continuous curves depending on the scaling factors.}
\label{currtyp}
\end{figure}
generation Fibonacci ring considering three different values of $N_e$.
The results are shown for both zero (Fig.~\ref{FD}(a)) and finite
(Fig.~\ref{FD}(b)) values of $\alpha$, where the solid, dashed and
dot-dashed lines correspond to $N_e=18$, $34$ and $56$, respectively.
For the ring without any SO coupling, currents are less fluctuating
with $N_e$, while the fluctuation becomes significant in the presence of
SO coupling. This is due to the irregular pattern of current amplitudes for
different current carrying states (Fig.~\ref{CCN}(e)). It is clearly 
observed from the spectrum Fig.~\ref{CCN}(e) that one or more states 
those carry smaller currents reside among the higher current carrying
states, and accordingly, when we set $N_e$ to a particular value, depending
on the top most filled energy level higher or smaller current is obtained
since the net current essentially depends on the contributions from the
neighboring states of this highest filled level.

\subsection{Anomalous oscillation of current with SO coupling}

The results analyzed so far are worked out only for some typical values 
of Rashba SO interaction. In order to establish the critical role played by
SO interaction more precisely on persistent current now we focus on the 
\begin{figure}[ht]
{\centering \resizebox*{6.25cm}{10cm}{\includegraphics{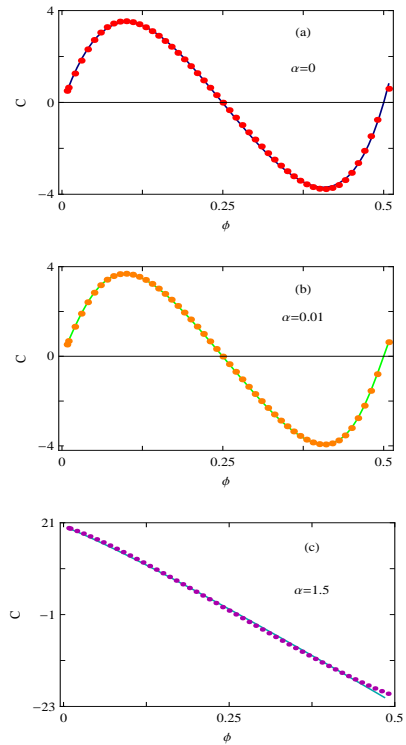}}\par}
\caption{(Color online). Dependence of $C$ on $\phi$ at three distinct 
values of $\alpha$. The dotted points are evaluated by exactly calculating
currents for a wide variation of ring size $N$ considering 
$\epsilon_A=-\epsilon_B=0.5\;$eV and $\beta=0$. Fitting these data sets we
generate functional forms which provide continuous curves.}
\label{SF}
\end{figure}
behavior given in Fig.~\ref{CCR}, where we plot persistent current as a
function of $\alpha$ for a particular flux $\phi=0.3\phi_0$. Both the
perfect and Fibonacci rings reflect the fact that typical current 
gets increased with increasing $\alpha$ providing anomalous oscillations.
Interestingly we see that in the impurity-free ring the typical 
current changes its sign alternately from positive to negative for 
a wide window of $\alpha$ and the window widths get broadened 
(Fig.~\ref{CCR}(a)) for higher values of $\alpha$. On the other hand, 
a continuous variation with smaller current (Fig.~\ref{CCR}(b)) is 
obtained in the Fibonacci ring. These features can be substantiated
from the spectra shown in Fig.~\ref{GER}. Here we plot the 
difference $\Delta E_0$ of ground state energies, determined at two 
typical fluxes ($\phi_{\mbox {\tiny typ}}$, $\phi_{\mbox {\tiny typ}}+
\Delta \phi$ ($\Delta \phi \rightarrow 0$)), as a function of $\alpha$
considering the same parameter values as taken in Fig.~\ref{CCR}. The
factor $-\Delta E_0/\Delta \phi$ gives the persistent current at 
$\phi_{\mbox {\tiny typ}}$, as used in conventional method, and thus from
the nature of $\Delta E_0$-$\alpha$ characteristics (Fig.~\ref{GER}) we
can estimate the oscillating behavior of current with $\alpha$ as
$\Delta \phi$ is always positive. This is exactly what we present in 
Fig.~\ref{CCR}.

\subsection{Scaling behavior}

Finally, in this sub-section, we discuss size-dependent persistent 
current in presence of SO interaction and from that we try to find 
the scaling behavior.

Figure~\ref{currtyp} demonstrates the variation of typical current 
$I_{\mbox {\tiny typ}}$ with system size $N$ for two different values of
$\alpha$ in the half-filled limit. Two cases are analyzed depending on
\begin{figure}[ht]
{\centering \resizebox*{6.5cm}{7cm}
{\includegraphics{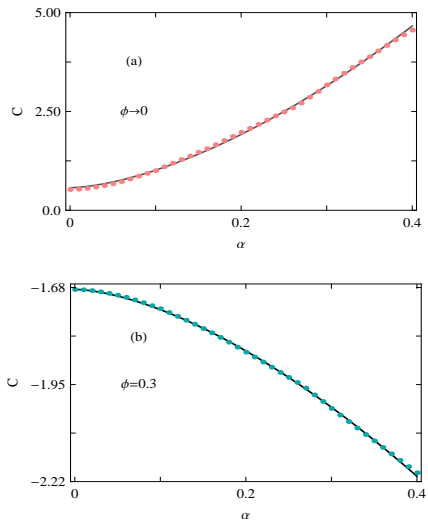}}\par}
\caption{(Color online). Dependence of $C$ on $\alpha$ for two different 
values of $\phi$. The colored dots and the continuous lines correspond
to the similar meaning as given in Fig.~\ref{SF}. The other parameters are:
$\epsilon_A=-\epsilon_B=0.5\;$eV and $\beta=0$.}
\label{SFalpha}
\end{figure}
the flux $\phi$, one is for $\phi \rightarrow 0$ while for the other we 
set $\phi=0.3$, and they are presented in (a) and (b), respectively.
The dots in the spectra are computed from the second-quantized approach
and they obey a scaling relation of the form: 
$I_{\mbox {\tiny typ}}=CN^{-\xi}$ where $\xi=1$ and $1.03$ for $\alpha=0$
and $1.5\;$eV, respectively, which we find from our extensive numerical
analysis. The pre-factor $C$ depends on both $\alpha$ and $\phi$. In 
the limit $\phi \rightarrow 0$, $C$ becomes $0.504$ and $19.718$ for
$\alpha=0$ and $1.5\;$eV, respectively, while these values are $-0.862$
and $-4.564$, respectively, for $\phi=0.3$. Using this scaling relation
we generate the continuous curves, where the black and red lines
correspond to $\alpha=1.5\;$eV and $0$, respectively. Clearly we see that
the curved lines fit the dots very well, and thus, we can utilize this
scaling relation to find charge current for any generation at these
typical values of $\alpha$ and $\phi$.

In analyzing Fig.~\ref{currtyp} two important aspects
should be noted. Firstly, the reduction of current with ring size $N$.
The reason behind this reduction can be easily understood in terms of 
the coherence of electronic wave function. For smaller rings wave 
function becomes coherent throughout the ring yielding larger current, 
while the phase coherence gets reduced with increasing $N$ providing 
lesser current. Secondly, the current amplitudes of different Fibonacci 
rings satisfy a specific scaling law. This scaling behavior essentially
comes from the quasiperiodicity of the system. The quasiperiodicity notably 
affects, as well, the energy band spectra of a Fibonacci ring (not shown 
here to save space). For instance, total energy bandwidth $\Delta E$ 
($=\sum_n |E_n(\phi=0)-E_n(\phi=\phi_0/2)|$, where $E_n$'s are the 
eigenvalues) sharply decreases with the Fibonacci generation $F_m$ and
it satisfies a similar kind of scaling relation with $F_m$. This is 
exactly reflected in $I_{\mbox {\tiny typ}}$-$N$ characteristics. 
Similar type of scaling 
nature is also obtained in other different quasiperiodic systems those 
have been described elsewhere~\cite{new1,new2,new3}.

Following this analysis one question naturally arises how the coefficient
$C$ depends on $\alpha$ and $\phi$, so that charge current can be estimated
\begin{figure}[ht]
{\centering \resizebox*{6.5cm}{6cm}{\includegraphics{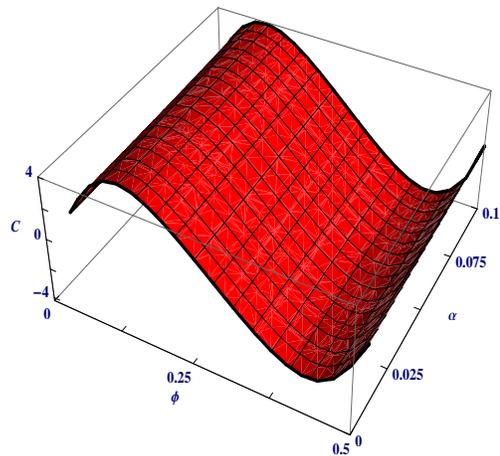}}\par}
\caption{(Color online). Simultaneous variation of $C$ with $\alpha$ 
and $\phi$ for the Fibonacci rings described with 
$\epsilon_A=-\epsilon_B=0.5\;$eV and $\beta=0$.}
\label{SFall}
\end{figure}
at arbitrary values of these parameters for any generation of the 
quasi-periodic Fibonacci ring. The answer is given in 
Figs.~\ref{SF}-\ref{SFall}. 

Focusing on the spectra given in Fig.~\ref{SF}, we see that for lower 
values of $\alpha$, $C$-$\phi$ data exhibits sinusoidal-like pattern,
though we cannot find a simple functional relationship with these data
sets, and accordingly, here we do not present that functional form. 
On the other hand, for $\alpha=1.5\;$eV, $C$-$\phi$ data can be fitted well
through a simple relation: $C=20.1-91 \phi^{1.1}$ and it gives a 
linear-like variation with $\phi$ (Fig.~\ref{SF}(c)).

In Fig.~\ref{SFalpha} we demonstrate $C$-$\alpha$ characteristics for two 
typical values of flux $\phi$. Two different functional forms are obtained 
for these fluxes and they are: $C(\alpha) = 0.57 + 17.72 \alpha^{1.6}$
(for $\phi\rightarrow 0$) and $C(\alpha) = -1.68 - 2.25 \alpha^{1.6}$
(for $\phi =0.3$).

At the same time, it is interesting as well as important to see 
the dependence of $C$ on both $\alpha$ and $\phi$ simultaneously. The 
result is given in Fig.~\ref{SFall} which clearly reflects the above 
scaling analysis as presented in Figs.~\ref{SF}-\ref{SFalpha}. 

\begin{figure}[ht]
{\centering \resizebox*{6.5cm}{7cm}{\includegraphics{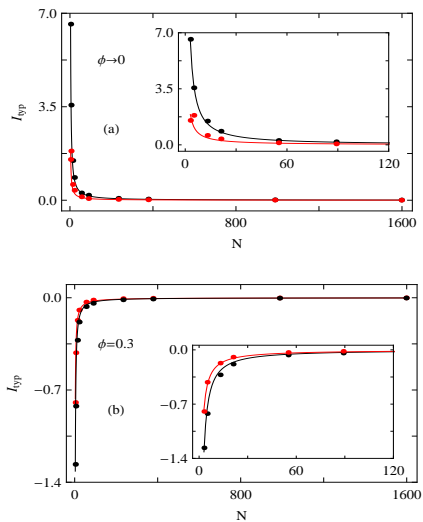}}\par}
\caption{(Color online). Variation of current with ring 
size $N$ considering $\epsilon_A=-\epsilon_B=0.5\;$eV and $\beta=0.5\;$eV 
where (a) $\phi \rightarrow 0$ and (b) $\phi=0.3$. The colored dots and the 
continuous lines correspond to the similar meaning as described in
Fig.~\ref{currtyp}.}
\label{newfig1}
\end{figure}

For the complete analysis of scaling behavior now we 
discuss the interplay of Rashba and Dresselhaus SOIs on persistent current. 
The results are shown in Fig.~\ref{newfig1} where the typical current
is calculated for two different values of Rashba SO coupling, like
Fig.~\ref{currtyp}, at two distinct AB fluxes ($\phi \rightarrow 0$
and $\phi=0.3$) considering the Dresselhaus SO coupling $\beta=0.5\;$eV.
The same scaling relation, viz, $I_{\mbox{\tiny typ}}=CN^{-\xi}$ is 
obtained where $\xi$ becomes $1$ and $1.08$ for $\alpha=0$ and $1.5\;$eV,
respectively. The pre-factor $C$ depends on both the SO couplings and
\begin{figure}[ht]
{\centering \resizebox*{6.5cm}{6cm}{\includegraphics{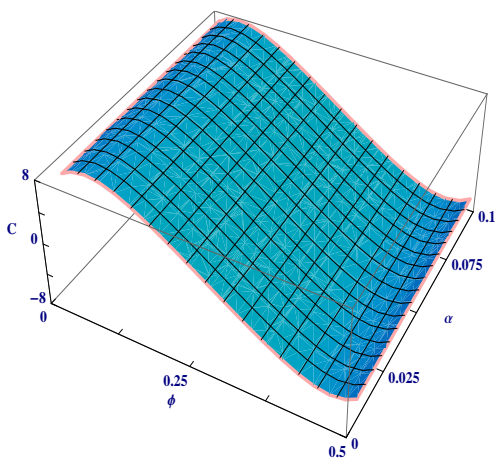}}\par}
\caption{(Color online). Simultaneous variation of $C$ 
with $\alpha$ and $\phi$ for the Fibonacci rings described with 
$\epsilon_A=-\epsilon_B=0.5\;$eV and $\beta=0.5\;$eV.}
\label{newfig2}
\end{figure}
flux $\phi$ as well. To have a complete idea about the 
variation of $C$ for a wide range of $\phi$ and $\alpha$ in presence of 
finite Dresselhaus SO coupling, in Fig.~\ref{newfig2} we present a $3$D 
diagram (like Fig.~\ref{SFall}), and, from this spectrum we can easily 
determine the pre-factor $C$ at the desired parameter values.

With these scaling results (Figs.~\ref{currtyp}-\ref{newfig2}) 
one can easily determine persistent current in any quasi-periodic Fibonacci 
ring in presence of SO couplings without doing detailed numerical 
calculations. Certainly this is a unique opportunity and has not been 
discussed before.
\vskip 0.3cm
\noindent
$\bullet$ {\bf Application Perspective:}
All the results described above are worked out only
for isolated rings. Now it is interesting and significant as well to 
know how such a system can be utilized in possible spintronic devices
since SO interaction in low-dimensional geometries has attracted
much attention due to its potential applications in diverse directions.
For example, to enhance quantum information processing as well as 
quantum computation controlling electron's spin degree of freedom is 
highly important~\cite{dev1}. The SO interaction can provide much deeper 
insight for generating spin current and also its manipulation rather 
than conventional methodologies. Sometimes the interplay between 
Rashba and Dresselhaus SO interactions may give a significant change 
in electronic transport, as discussed by several groups~\cite{dev2,dev3}. 
In order to reveal these facts the ring has to be connected with external
electrodes, viz, source and drain. In presence of such electrodes one
can study different aspects of spin-dependent transport like, spin
currents, spin resoled conductances, spin polarization to name a few.
One such work has been done Wang and Chang towards this 
direction~\cite{dev4}. They have studied two-terminal spin-dependent
transport through a 1D ring subjected to both Rashba and Dresselhaus
SO couplings. The interplay between these two SOIs leads to a significant
change in electronic transmission, localization of electrons and also
spin polarization of the current. They have also shown how conductance
is sensitive to the ring-electrode interface geometry, and these results
certainly give a great impact in designing future spintronic devices.
Several other works~\cite{mlt1,mlt2,mlt3,mlt4,mlt5} have also been put 
forward along this direction to explore many interesting features 
of spin transport in a bridge setup. Before the end, we would like to
state that since the study in open system (viz, source-ring-drain system) 
requires a complete separate theoretical approach, here in the present 
manuscript we do not go for this. We will analyze these aspects in our 
future work.

\section{Closing Remarks}

In conclusion, we have investigated the critical roles played by Rashba
and Dresselhaus SO couplings on persistent charge current in a 
quasi-periodic Fibonacci ring threaded by a magnetic flux $\phi$. Using
a tight-binding framework we have computed individual state currents as
well as net current for a particular band filling based on second-quantized
approach. Analyzing state currents we can predict the conducting nature
of individual energy levels, which on the other hand, provides an important
tool in understanding the net response of a complete system. From the
calculation of net current we have found that SO interaction can enhance
the current significantly and sometimes it becomes orders of magnitude higher
compared to the SOI-free Fibonacci ring. This observation 
might throw some light in the era of deep-rooted doubt between the 
experimental observations and theoretical predictions of persistent current.

In the rest of our work, we have essentially focused on the scaling
behavior of persistent current with ring size $N$, associated with the
Fibonacci generation, and established a unique way of determining persistent
charge current without going through detailed numerical calculations. 
In the analysis of scaling properties we have restricted ourselves to
the half-filled band limit considering odd electron filling. But, these 
scaling relations can be well applied to the even electron filling
for the half-filled band case, expect the small rings (viz, $F_5$ and 
$F_8$) where
the current deviates slightly from our fitting curve. Indeed, the
establishment of scaling relation for any general electron filling and 
for any disordered ring, be it random or made of any kind of quasi-periodic
lattices, will be highly interesting and important too. These issues 
will be available in our next work and it is the first step towards this
direction. 

Finally, it should be important to note that throughout the analysis we
have presented the results only for the site model. But almost identical 
features are also obtained for the bond model and even for the mixed
model, which we verify through our exhaustive numerical analysis, and
accordingly, here we do not present those results.

\end{document}